\documentstyle[aps,preprint,12pt]{revtex}

\begin{document}

\title{ Stochastic Coherence in Coupled Map Lattices }
\author{ Manojit Roy$^a$ and R. E. Amritkar$^b$ \\
\it $^a$Department of Physics, University of Pune, Pune 411007, India \\
\it $^b$Physical Research Laboratory, Navrangpura, Ahmedabad 380009,
India }
\maketitle

\begin{abstract}
We report in details the observations of {\it structures} 
in coupled map lattice during its chaotic evolution,
both in one and two dimension,
driven by identical noise on each site (by a structure we mean a group of
neighboring lattice--sites for whom values of dynamical variable follow
certain prespecified pattern). It is observed that the number of these
structures decays with their size following a {\it power--law}, for a given
noise--strength. The number of structures decays with their lifetime
following a stretched exponential.
We have seen an interesting phenomenon, which we call
{\it stochastic coherence}, in which the average length as well as the
average lifetime
of these structures exhibit {\it bell--shaped} maxima for some intermediate
range of noise--strength values. Similar features have also been observed
for `spatio--temporal structures'.
\end{abstract}

\vspace{.5in}

\hspace{.4in} {\bf Keywords:} Chaos; noise; coupled map lattice.


\section{Introduction}

Over years coupled map lattice (CML) has become a very popular model to
study a host of physical phenomena involving systems with spatial
dimension~\cite{OP,OK,GA,AG,KK,YK,ZG,RG,SBAL,KF,AGS,CG,JA}. In particular,
CML exhibits a variety of spatio--temporal patterns and
structures and has been extensively used, with varying degree of success,
to model similar features in experimental systems such as
Rayleigh--B\'enard convection, Taylor--Couette flow, B--Z reaction
etc.~\cite{OP,OK,KK,YK,CH}. One of the challenging problems is to 
understand
the formation of structures, localized in both space and time, in turbulent
fluid~\cite{AKMH,HKM,JCRH}. Recently lot of attention has been devoted to
the role of fluctuations in onset, selection and evolution of such patterns
and structures~\cite{MM,HS,BPTK,WK,FH,DAK,MB,JMK,GB}. An interesting and
counterintuitive observation is that the presence of noise can help sustain  
structures which otherwise would have been absent in an evolving dynamical
system.

In this paper we present the detailed account of a novel phenomenon that we
have recently observed~\cite{RA} in the dynamics of structures in a
chaotically evolving
CML, in one and two dimension, driven by identical noise.
We call a {\it structure} as a group of neighboring sites whose
variable--values
follow certain prespecified spatial pattern. As the lattice evolves under
the influence of noise, the number of these structures exhibits
{\it power--law decay} with size of the structure
for a given noise--strength, with an exponent which is a function of 
noise--strength. It is observed that average size of these structures,
plotted against noise--strength, shows a bell--shaped curve with a
characteristic peak. Similar features have been noted for the
`spatio--temporal structures' also (by this we mean the cluster of sites
which remain in a structure as it evolves in time). Average lifetime 
of these structures also exhibit a maximum within same range of noise values.
The behaviors remain essentially the same in two dimensional CML as well.
We call this new phenomenon {\it stochastic coherence}.

\section{The system}

We first consider a one dimensional CML with the dynamics
\begin{eqnarray}
x_{t+1}(i) = (1 - \varepsilon) F(x_t(i)) + {\varepsilon \over 2} 
\Big[ F(x_t(i-1)) + F(x_t(i+1))  \Big] + \eta_t\;, \label{sc1} 
\end{eqnarray}
where $x_t(i)$, $i = 1, 2, \cdots, L$, is value of the variable located
at site $i$ at time $t$, $\eta_t$ is the additive noise, $\varepsilon$ is
the (nearest neighbor) coupling strength, and $L$ is the size of the
lattice. Evolution on the lattice sites is governed by the nonlinear
logistic function
\begin{equation}
F(x) = \mu x (1 - x), \label{logistic}
\end{equation}
where $\mu$ is the nonlinearity 
parameter. Both open--boundary conditions 
$x_t(0) = x_t(1),\;x_t(L+1) = x_t(L)$,
and periodic--boundary conditions $x_t(L+i) = x_t(i)$, have been used for
our system. For noise $\eta_t$ we have chosen a uniformly distributed random
number bounded between $-W$ and $+W$. We call $W$ the {\it noise--strength}
parameter.

We define a {\it structure} as a region of space such that the dynamical
variables at sites within this region follow a prespecified spatial
pattern~\cite{JA}. For our purpose we choose a spatial pattern where the
difference in the values of the variables of neighboring sites within the
structure is less than a predefined small positive number say $\delta$, i.e.,
\begin{eqnarray}
\vert x_t(i) - x_t(i\pm1) \vert \leq \delta\;. \nonumber
\end{eqnarray}
We call $\delta$ the {\it structure parameter}. We study the distributions
of abundance and lifetimes of these coherent structures as the CML
evolves chaotically.

\section{Results and discussion}

Values of the system parameters $\mu$, $\varepsilon$ and $L$ are so
chosen that the dynamics of the system remains essentially chaotic. 
We have taken typical values $L = 10000,\:\mu = 4$ and $\varepsilon = 0.6$.
Value of structure parameter $\delta$ is chosen as $\delta = 0.0001$.
Coherent
structures with length $<3$ (sites) and lifetime $<2$ (timesteps) are
disregarded. $W$ is restricted to within $[0,1]$. For the
results presented here open--boundary conditions are used.
80000 transient steps have been discarded and the
data are integrated over 100000 iterates per initial condition and 4 
different initial conditions.

\subsection{Distribution of $n(l)$ vs. $l$}

We have investigated the distribution of number $n(l)$ of structures
with their length $l$, for different values of noise--strength
parameter $W$.
In Fig.~1 we show the plot (on log--log scale) of $n(l)$ against $l$, 
for five values of $W$. The plot exhibits, over a wide range of scales,
a characteristic {\it power--law} of the form
\begin{eqnarray}
n(l) \propto l^{-\alpha_1}\;, \label{sc3}
\end{eqnarray}
where $\alpha_1$ is power--law exponent.

If one uses only a uniform--deviate random number as site--variable,
probability that a site belongs to a structure of length $l$ is 
$p_\delta(l) \approx l(2\delta)^{l-1}(1 - 2\delta)^2$.
So the number of such structures in a lattice of size $L$ will be
\begin{eqnarray}
n_\delta(l) \approx L(2\delta)^{l-1}(1 - 2\delta)^2\;, \label{sc4}
\end{eqnarray}
which shows an exponential decay. Thus correlations in the system are
important to get a power law form.

\subsection{Stochastic coherence}

One can see from Fig.~1 that the exponent $\alpha_1$ of 
relation~(\ref{sc3}) depends on noise--strength $W$. Fig.~2 shows the
variation of $\alpha_1$ with $W$. $\alpha_1$ is exhibiting a minimum for 
values of $W$ around $0.6$. To get a clearer picture, let us 
define average length $\bar{l}$ of a structure as
\begin{eqnarray} 
\bar{l} = \sum{l\,n(l)} / \sum{n(l)}. \label{sc5}
\end{eqnarray}
We have investigated how noise influences this quantity.
In Fig.~3 we plot the variation of $\bar{l}$ with $W$ for values of 
parameters as in Fig.~1. One sees a quite interesting {\it bell--shaped} 
dependence for $W$ around value $0.6$.

This behavior may remind one of the much studied stochastic resonance 
phenomenon~\cite{BPSV,MWR,FL}, a signature of which is the existence of
a bell--shaped behavior of the temporal response of the
system, when plotted against noise--strength. The similarity, however,
is purely coincidental because there exists one important difference. 
In stochastic resonance noise transfers energy to the system at a given
time scale, which usually is the characteristic scale of the system. On
the other hand, power law feature of Fig.~1 implies that our system
does not have any preferred length scale; noise {\it induces coherence} 
on all length scales. We call this novel phenomenon {\it stochastic
coherence}.

\subsection{Stability analysis}

Let us see, qualitatively, how noise modifies the stability matrix 
of the system, thereby influencing its stability properties. We 
consider stability matrix $M$ of a homogeneous state 
\{$\cdots,x_t,$ $x_t,x_t,\cdots$\} at time $t$. This state may be 
imagined
as a large structure with $\delta = 0$. One time step later the 
stability matrix is
\begin{eqnarray}
M_{t+1} = JF_t^\prime\;, \label{sc6}
\end{eqnarray}
where $J$ is the familiar tridiagonal matrix
\begin{equation}
J=\left\{
\begin{array}{ccccc}
\ddots & \ddots \\
\ddots & 1-\varepsilon & \varepsilon/2 & 0 &  \\
 & \varepsilon/2 & 1-\varepsilon & \varepsilon/2 & \\
 & 0 & \varepsilon/2 & 1-\varepsilon & \ddots \\
 &  &  &  \ddots & \ddots 
\end{array}
\right\}
\end{equation}
with $1 - \varepsilon$ as its diagonal elements and $\varepsilon/2$ as
offdiagonal elements on either sides, and 
\begin{equation}
F_t^\prime \equiv F^\prime (x_t) (\equiv {dF \over dx}) = \mu (1 - 2 x_t)\;. 
\nonumber
\end{equation}
After two more timesteps, the stability matrix
takes the form
\begin{eqnarray}
M_{t+1}M_{t+2}M_{t+3} & = & J^3 F_t^\prime F_{t+1}^\prime F_{t+2}^\prime 
\nonumber \\
& = & J^3F_t^\prime \mu^2 [(1 - 2F_t)\{1 - 2 \mu F_t (1 - F_t)\}
+ 6 \mu \eta_t^2(1 - 2F_t)
\nonumber \\
& & - 2 \eta_t\{1 + \mu (1 - 6F_t + 6F_t^2)\} -
4 \mu \eta_t^3 - 2 \eta_{t+1}(1 - 2F_t) + 4 \eta_t \eta_{t+1} ]\;.
\nonumber
\end{eqnarray}
We now 
average this expression over the noise distribution. The noise has uniform
distribution, with zero mean and delta--correlation. By averageing, the
terms containing cross terms in $\eta$ (i.e. terms of the type
$\eta_t \eta_{t+1}$) and odd powers in $\eta$ vanish,
leaving the following expression with only even powers in
$\eta$:
\begin{eqnarray}
<M_{t+1}M_{t+2}M_{t+3}>\;=\;J^3 (F_t^\prime)^2 \mu \Big[1 - 2 \mu F_t (1 -
F_t) + 6 \mu <\!\eta_t^2\!> \Big], \label{sc7}
\end{eqnarray}
where $<>$ denotes averaging over noise--distribution. One more averaging
is needed, this time over the invariant distribution of the CML. Fig.~4
shows the invariant density of a single site of CML,
with $W = 0.6$ and the other system
parameters same as mentioned in the beginning of this section.
The density is highly asymmetric, with larger weightage for $x(i) > 0.5$.
Because of this asymmetry, averaging expression (\ref{sc7}) over the 
invariant density makes the term $\;1 - 2 \mu F_t (1 - F_t)\;$ negative. 
This,
added to the positive noise term $\;6\mu<\eta_t^2>$, results in reduction 
of eigenvalue
of the above matrix. This in turn implies a consequent reduction of the
instability of the state (or the structure). Thus noise plays a crucial
role in enhancing the abundance of the structures.

\subsection{Evolution of structures}

Let us now consider the evolutionary aspects of these coherent structurs.
We have studied the distribution of the number $n(\tau)$ of structures 
having lifetime $\tau$. In order to obtain $\tau$, each structure is 
tracked as the lattice evolves, till the structure degenerates completely.
In Fig.~5 we show such an evolutionary diagram for a CML with a typical
size 
$L = 100$, evolving chaotically for 200 timesteps. The shaded regions 
depict the evolution of coherent structures. Fig.~6 shows the variation,
on log--linear scale, of $n(\tau)$ with $\tau$, for different values of
noise $W$. Number $n(\tau)$ exhibits a decay with $\tau$ with a {\it 
stretched exponential} form
\begin{eqnarray}
n(\tau) \propto \exp\big(-({\rm const.})\tau^\beta\big)\;. \label{sc8}
\end{eqnarray}
The exponent $\beta$ is seen to depend on $W$. Like in case of $\alpha$ of
expression (\ref{sc3}), $\beta$ also exhibits a minimum near $W = 0.6$.

We define average lifetime $\bar{\tau}$ of a structure as 
\begin{eqnarray}
\bar{\tau} = \sum{\tau\,n(\tau)} / \sum{n(\tau)}\;, \label{sc9}
\end{eqnarray}
and explore its dependence on noise strength $W$.
Fig.~7 shows a plot of $\bar{\tau}$ vs. $W$. It again exhibits a 
maximum for $W$ around $0.6$.

There is another feature of the system which merits an elaborate study. 
This concerns the distribution of `spatio--temporal structures (STS)'.
By such
a structure we mean the group of sites that remain in a coherent structure
as it evolves in time, till it completely degenerates. Each shaded region
of Fig.~5 constitutes one such STS. In Fig.~8 we plot
the number $n(S_1)$ of an STS vs. its `size' $S_1$ for different noise 
values $W$. We again see an interesting
power law dependence of the type (\ref{sc3}) for an appreciable range of
sizes, but with different exponent.
This exponent also depends on $W$, with a minimum around $W=0.6$. Let us
define average size $\bar{S_1}$ of an STS in a similar fashion as 
\begin{eqnarray}
\bar{S_1} = \sum{S_1\,n(S_1)} / \sum{n(S_1)}. \label{sc10}
\end{eqnarray}
We plot $\bar{S_1}$ as function of $W$ in 
Fig.~9. This plot also exhibits a characteristic bell shape within a range
around $W=0.6$, quite similar to the plot in Fig.~3.

\subsection{Lyapunov spectrum analysis}

We have studied the lyapunov exponent spectrum ($\lambda$ spectrum) of
the CML, and its dependence on noise $W$ as well as on the coupling
parameter $\varepsilon$. We find that a number of lyapunov exponents is 
positive, thereby implying that the system is indeed evolving chaotically. 
In Fig.~10 we plot the variation of maximum exponent $\lambda_{\rm max}$
with noise strength $W$, for coupling parameter $\varepsilon = 0.6$.
$\lambda_{\rm max}$ exhibits a clear minimum around $W=0.6$~\cite{MT},
the same region where all the extrema reported above are located. 
This observation may create the impression that the reduction of lyapunov
exponent alone is responsible for stochastic coherence.

To explore this possibility, we have studied the dependence of both
$\lambda_{\rm max}$ and average length $\bar{l}$ on coupling strength
$\varepsilon$. Fig.~11 plots the variation of $\lambda_{\rm max}$ with
$\varepsilon$ for different $W$. It is clear that
$\lambda_{\rm max}$ remains fairly constant for $0.2 \leq \varepsilon 
\leq 0.8$ for all $W$. On the other hand, $\bar{l}$ shows a monotonic 
increase with $\varepsilon$ for all $W$. In Fig.~12 we show one such plot 
of $\bar{l}$ against $\varepsilon$, for $W = 0.6$.
These two contrary features indicate that lyapunov exponent alone
is not good enough to characterize the full range of spatio--temporal 
features of the system.

It is worth pointing out that stochastic coherence is quite distinctive
phenomenon as compared to spatio--temporal intermittency~\cite{CM}. 
In intermittency the system exhibits alternately regular and chaotic 
bursts in time. Our system does not show any such behavior. This fact is
further corroborated by the fact that the power spectrum of time--series 
does not have any peak for entire range of noise--strength. Thus our 
system is undergoing essentially a 
{\it spatially intermittent and temporally chaotic} evolution.

\subsection{Full lattice coherence (FLC)}

One quite interesting phenomenon that we have observed is that
for $W \geq 0.4$ often the entire lattice itself evolves as a
single coherent structure. This state should be distinguished from the
synchronized state where at any given instant $t$, $\;x_t(i) = x_t$ for
$i = 1, 2, \cdots, L$ (that is, $\delta = 0$). On occasions this {\it full 
lattice coherence} (FLC)
survives for quite long durations (as long as 200 timesteps or more), but
eventually degenerates completely. This implies that synchronized
state is not a stable attractor for our system. We have confirmed this by
calculating the largest Lyapunov exponent for the synchronized state
which turns out to be positive.

However, occasionally our
system did fall into an apparently synchronized state after a very large
time (larger than $10^7$ steps). This happens because of finite accuracy of
computation which cannot differentiate between unstable and stable
synchronized states~\cite{ASP}. Let us investigate this point further.
We define a quantity $\bar{T}$ as the average time required for first
occurrence of FLC, the averaging taken over different initial conditions.
In Fig.~13 we plot, on log--log scale, $\bar{T}$ against the structure 
parameter $\delta$, for a fixed lattice size $L = 10000$. The plot 
exhibits a power law of the form
\begin{eqnarray}
\bar{T}_L \propto \delta^{-\gamma}\;. \label{sc11}
\end{eqnarray}
This relation implies that the synchronized state ($\delta = 0$) will not
occur in a finite lattice. We now show, in Fig.~14, the plot of $\bar{T}$
against lattice size $L$, for a fixed $\delta$. This is again a power law
of the type
\begin{eqnarray}
\bar{T}_\delta \propto L^\nu\;. \label{sc12}
\end{eqnarray}
We thus see that as $L \rightarrow \infty$, even FLC with a nonzero
$\delta$ will not be attained. Therefore, relations (\ref{sc11}) and
(\ref{sc12}) tell us that synchronized state for our system is an
artifact, resuting due to the combined effect of finite lattice size and
finite computational accuracy.

Existence of FLC in a finite lattice can be seen to be
a consequence of the power--law (\ref{sc3}). Probability of a site to
belong to a structure of length $l$, following (\ref{sc3}), is 
$p_\delta(l) \propto l.l^{-\alpha_1} = l^{1-\alpha_1}$. The probability 
of a
site to belong to a structure of length $\geq L$ is then $P_0
\equiv P_\delta(\geq L) \propto \sum_L^\infty{l^{1-\alpha_1}} \approx
\int_L^\infty{dl\,l^{1-\alpha_1}} \propto L^{2-\alpha_1}$ (this holds for 
$\alpha_1 > 2$, which is true for our system as can be seen from Fig.~2). 
Thus, the probability that for
the first time a site belongs to a structure of length $\geq L$ at
timestep $T$ is
$P_\delta(T) \propto (1 - P_0)^{T-1}P_0$. Average of $T$ is
$\bar{T}_\delta(\geq L) = \sum_{T=1}^\infty{T\,P_\delta(T)} \propto
(P_0)^{-1}$, i.e.,
\begin{eqnarray}
\bar{T}_\delta(\geq L) \propto L^{\alpha_1-2}\;. \label{sc13}
\end{eqnarray}
This shows that FLC (with nonzero $\delta$) will occur in a finite 
lattice. From Fig.~2 we get $\alpha_1 - 2 \approx 0.22$ for $W = 0.6$.
On the other hand, we estimate (from Fig.~14) $\nu$ of relation 
(\ref{sc12}) to be approximately 0.3. The discrepancy may occur due to
the fact that relation (\ref{sc13}) holds for an infinite lattice,
whereas relation (\ref{sc12}) is numerically obtained for finite
lattices. 

\section{Two dimensional CML}

Let us now consider a two dimensional CML. The dynamics of this system
takes the form
\begin{eqnarray}
x_{t+1}(i,j) = (1 - \varepsilon) F(x_t(i,j)) + {\varepsilon \over 4}
 & \Big[ & F(x_t(i-1,j)) + F(x_t(i+1,j)) \nonumber \\ &&
+ F(x_t(i,j-1)) + F(x_t(i,j+1)) \Big] + \eta_t\;, \label{sc14}
\end{eqnarray}
where $x_t(i,j)$, $i,j = 1, 2, \cdots, L$, is the value of the variable at
site ($i,j$) in the lattice having size $L \times L$, and $F(x(i,j)) = 
\mu\;x(i,j) (1 - x(i,j))$. For structures in 
this lattice we look for those sites ($i,j$) with 
\begin{eqnarray}
\vert x_t(i,j) - x_t(i\pm1,j) \vert \leq \delta\;, \nonumber
\end{eqnarray}
and, 
\begin{eqnarray}
\vert x_t(i,j) - x_t(i,j\pm1) \vert \leq \delta\;. \nonumber
\end{eqnarray}
As the CML evolves, we study the distributions of abundance and lifetimes
of these two dimensional structures, as we have done for one dimensional
system.

We present some of the observations in the following. Again we have chosen
the system parameters so that the CML evolves chaotically. Typical values 
taken are $L \times L = 100 \times 100$, $\mu = 4$, and $\varepsilon = 
0.6$. We have chosen structure parameter $\delta = 0.0001$. 
3000 transient steps have been discarded, and data are obtained
over 20000 iterates per initial condition and 4 initial conditions. We
have used open boundary conditions for the results presented here.

Fig.~15
exhibits the plot (on log--log scale) of distribution of the number $n(s)$
of coherent structures against their size $s$ for different $W$ (we use
the notation $s$ to denote size of the structures in two dimensional CML,
unlike $l$ which has been used for one dimensional case. This is also to
be distinguished from $S_1$, the notation used for sizes of STS). 
One can see a power law of the form
\begin{eqnarray}
n(s) \propto s^{-\alpha_2}\;, \label{sc15}
\end{eqnarray}
similar to the relation~(\ref{sc3}), for a wide
range of values of $s$ (for $s\geq 10$; there is a slight bend in the
graphs for $s \leq 10$). 
The exponent $\alpha_2$ for the power law
is slightly less in value than its one dimensional counterpart 
$\alpha_1$, for all
$W$, indicating a less steeper decay. These exponents again
show a dip around $W = 0.6$. We define the average size $\bar{s}$ of the 
structures as 
\begin{eqnarray}
\bar{s} = \sum{s\,n(s)} / \sum{n(s)}\;. \label{sc16}
\end{eqnarray}
$\bar{s}$
exhibits a corresponding peak around $W = 0.6$. This can be clearly
seen in the bell--shaped curve of Fig.~16, which is the plot of $\bar{s}$
against $W$. This plot is quite alike the one of Fig.~3.

In Fig.~17 we plot, on log--linear scale, distribution of number $n(\tau)$
of structures against their lifetime $\tau$. The plot shows a stretched
exponential type behavior of the form (\ref{sc8}), similar to Fig.~6.
Fig.~18 shows the variation of average lifetime $\bar{\tau}$, as defined in
(\ref{sc9}), with noise strength $W$. We see a peak near $W = 0.6$, quite
similar to the plot of Fig.~7.

Distribution of number $n(S_2)$ of STS is plotted, on log--log scale, 
against size $S_2$ in Fig.~19. We see a very similar power law decay as
in Fig.~8 for one dimensional CML. As in previous instances, we define
average size $\bar{S_2}$ of STS as
\begin{eqnarray}
\bar{S_2} = \sum{S_2\,n(S_2)} / \sum{n(S_2)}. \label{sc17}
\end{eqnarray}
Fig.~20 shows the variation of $\bar{S_2}$ with noise strength $W$. The
plot shows a bell--shaped peak around $W = 0.6$, similar to the plot of
Fig.~9.

All the other features that are observed
in one dimensional system are also seen in two dimensional CML. For 
instance, we have seen full lattice coherence for $W \geq 4$, which persist
for considerable duration before eventually breaking up.

We have repeated all the above observations, for both one and two
dimensional systems, for several other values of the coupling parameter
$\varepsilon$ ranging from $0.1$ to $0.9$, as well as for nonlinearity
parameter $\mu$ between $3.6$ and $4$. We find that all the features
remain essentially the same, indicating the robustness of the 
phenomenon. We have also observed similar results using periodic--boundary
conditions for the lattice.

\section{Conclusion}

We have investigated in details a new phenomenon associated with structures
in a chaotically evolving CML, in both one and two
dimensions, driven by identical noise. We call this phenomenon
{\it stochastic coherence}. We have seen a marked rise in abundance of
coherent structures of all scales due to noise. Stability matrix of these
structures shows that noise can reduce their instability, thereby
enhancing abundance.. Distribution of
the structures exhibits a power--law decay with size of the structure,
with an exponent having a minimum at some intermediate noise--strength. 
Average size of these structures shows a bell--shaped maximum at the same
value of noise--strength. This feature is similar to that of the stochastic
resonance phenomenon. However, our system does not have any intrinsic
length--scale, whereas stochastic resonance is associated with a given
time--scale. We have also observed similar maxima for the average lifetime
of these structures as well as for the average size of `spatio--temporal
structures', at the value of noise close to the earlier extrema. Our
phenomena may have their importance in understanding the interesting role
that noise plays in formation and evolution of structures in spatially
extended systems.

\section{Acknowledgements}
One of the authors (MR) acknowledges University Grants Commission (India)
and the other (REA) acknowledges Department of Science and
Technology (India) for financial assistance.

\newpage
\begin{center}

{\bf FIGURE CAPTIONS}

\end{center}
\vspace{0.2 in}

\begin{itemize}
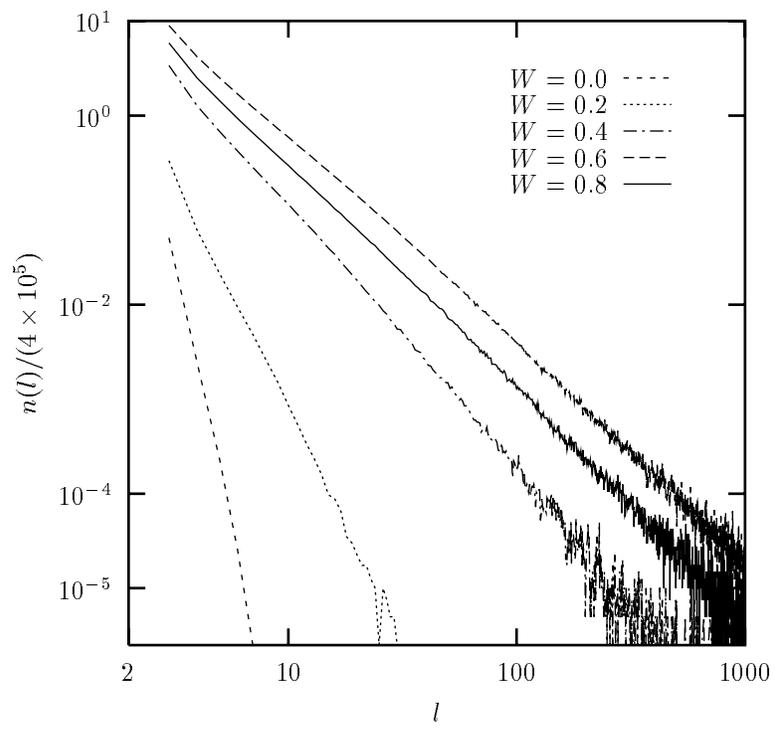
\item[Fig.~1.] The figure shows the variation, on log--log scale, of number
 $n(l)$ of structures with length $l$ for a one dimensional lattice
with size $L = 10000$,
for different values of noise--strength $W$ as indicated. The system
parameters are coupling strength parameter $\varepsilon = 0.6$, structure
parameter $\delta = 0.0001$, and nonlinearity parameter $\mu = 4$.
Open--boundary conditions are used. 80000 transient steps are discarded.
Data are obtained for 100000 iterates per initial condition and with 4
initial conditions.

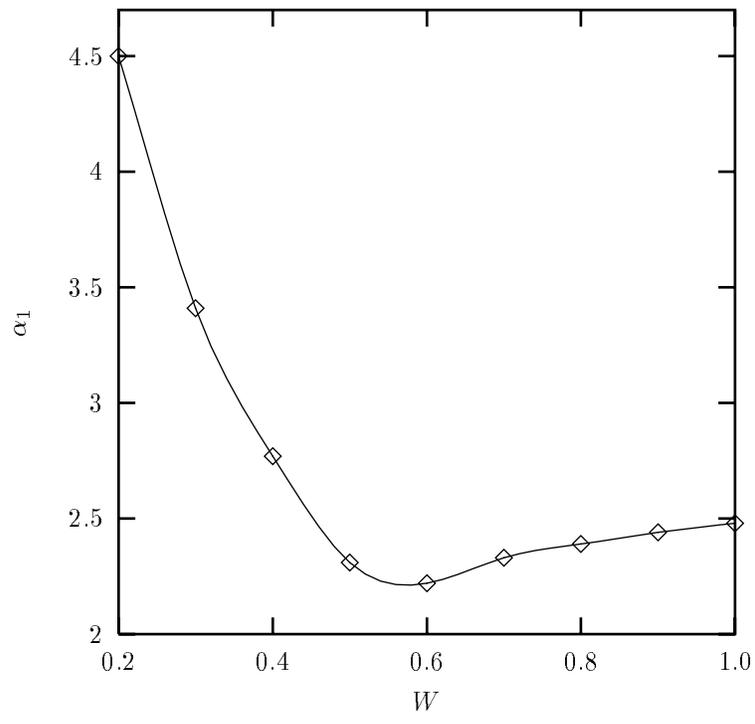
\item[Fig.~2.] Exponent $\alpha_1$ of the power law (\ref{sc3}) (slope of
the plots in Fig.~1) is plotted against noise strength $W$, with 
system parameters as in Fig.~1.

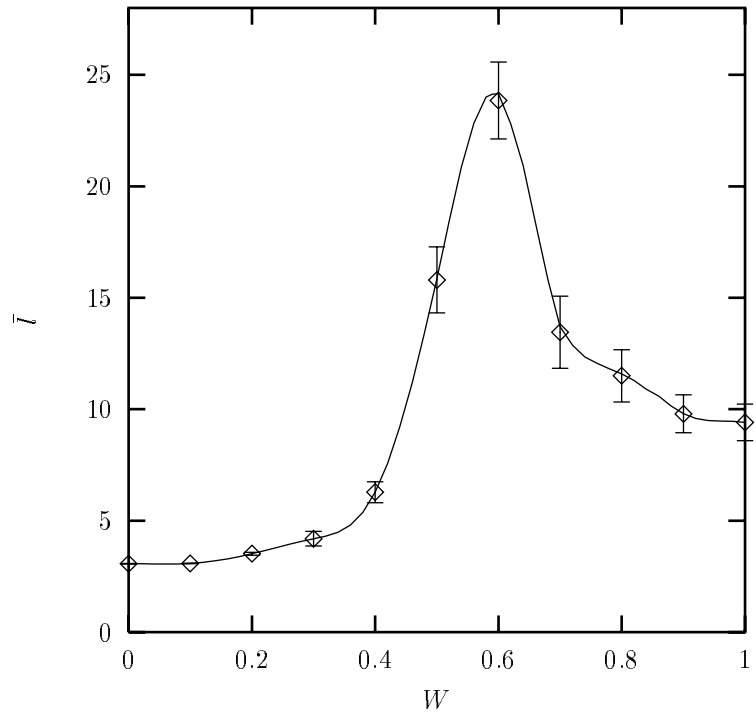
\item[Fig.~3.] Variation of average length $\bar{l}$ of structure with 
$W$ is plotted, with parameters as stated in Fig.~1.
Vertical error bars indicate the standard deviations at the data points.

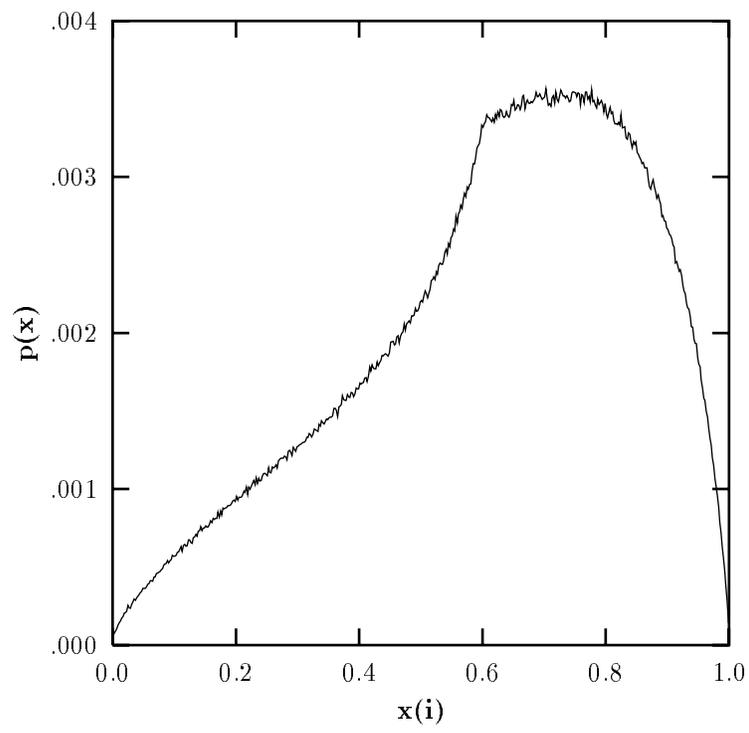
\item[Fig.~4.] Invariant density $p(x)$ of the CML is plotted against
variable value $x$, for the system
parameters as for Fig.~1 and with noise value $W = 0.6$. The
asymmetry is clearly evident, with a much higher weightage for $x > 0.5$.

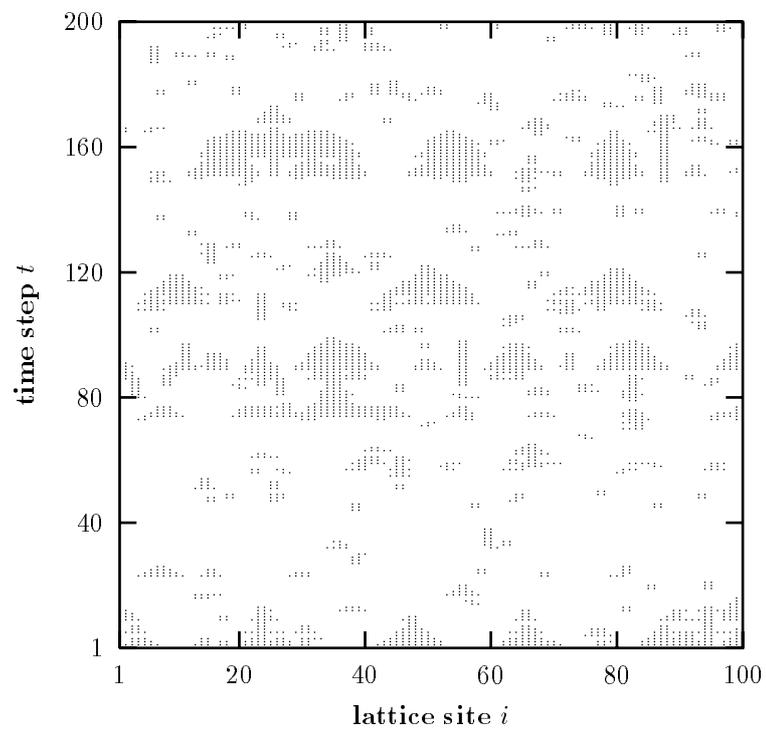
\item[Fig.~5.] Evolution of coherent structures is shown by shaded
regions, during 200 time steps, in a CML with a typical size $L = 100$, 
and with noise value $W = 0.6$, and other parameters same as in earlier 
figures. 

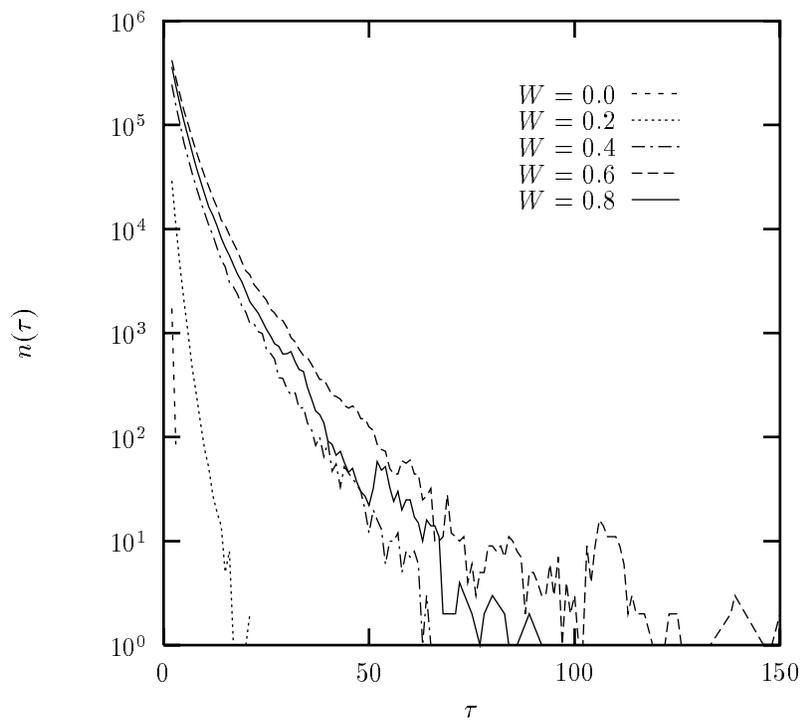
\item[Fig.~6.] Distribution of number $n(\tau)$ of the structures is
plotted, on log--linear scale, against their lifetime $\tau$. The system
parameters are as in Fig.~1.

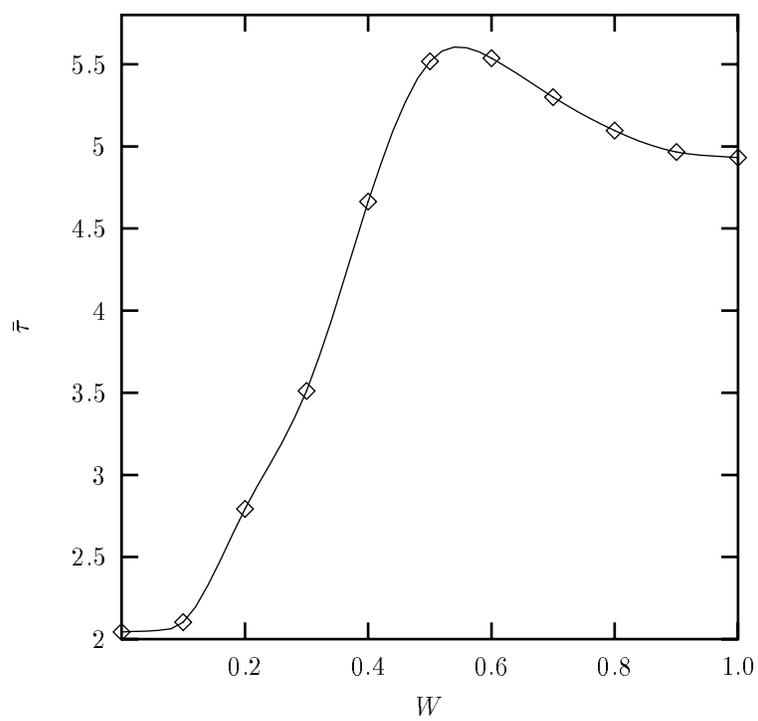
\item[Fig.~7.] Plot shows the variation of average lifetime $\bar{\tau}$
of structures with noise--strength $W$, for parameters as in Fig. 1.

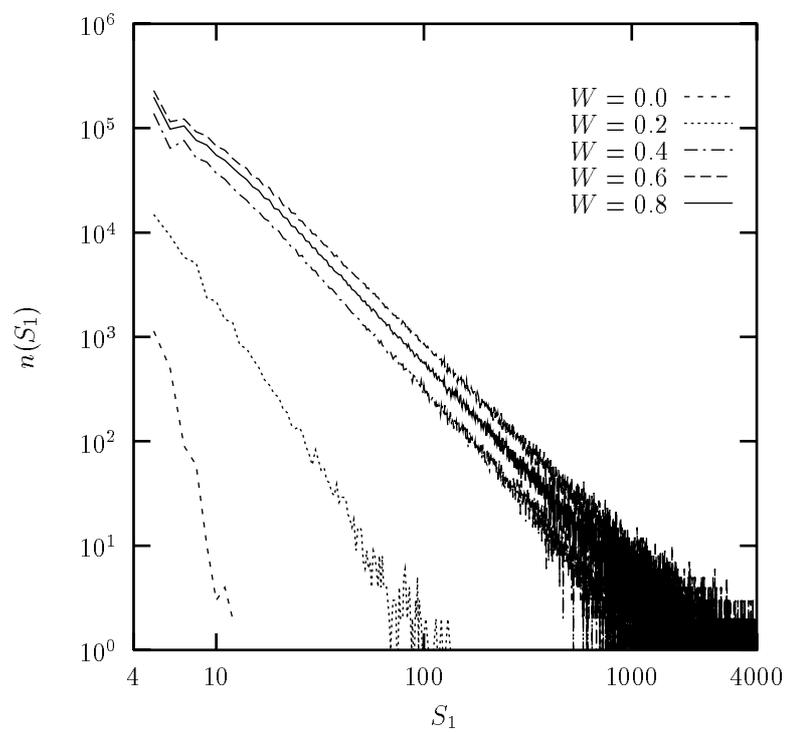
\item[Fig.~8.] Distribution of the number $n(S_1)$ of spatio--temporal 
structures (STS) is plotted, on log--log scale, against their size $S_1$, 
for the same parameter values as in Fig. 1.

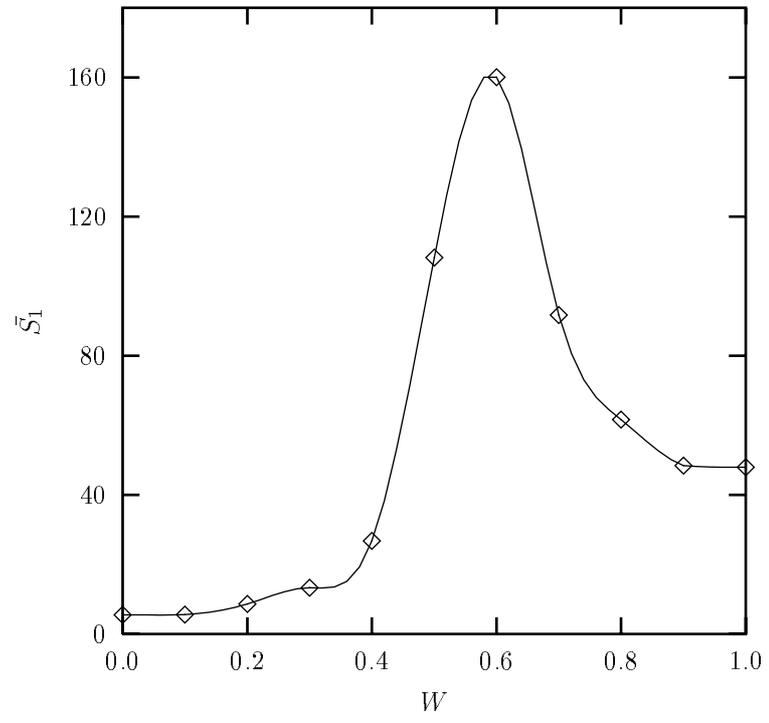
\item[Fig.~9.] The plot shows the variation of average size $\bar{S_1}$ of
STS with the noise--strength $W$, for parameter values as in earlier
figures.

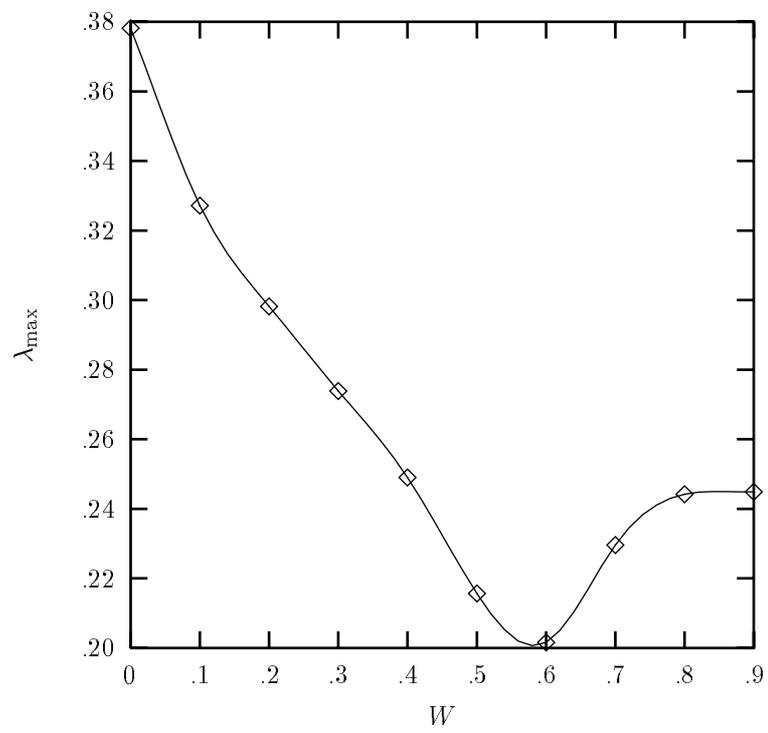
\item[Fig.~10.] Variation of $\lambda_{\rm max}$ is plotted against noise
strength $W$, for $\varepsilon = 0.6$.

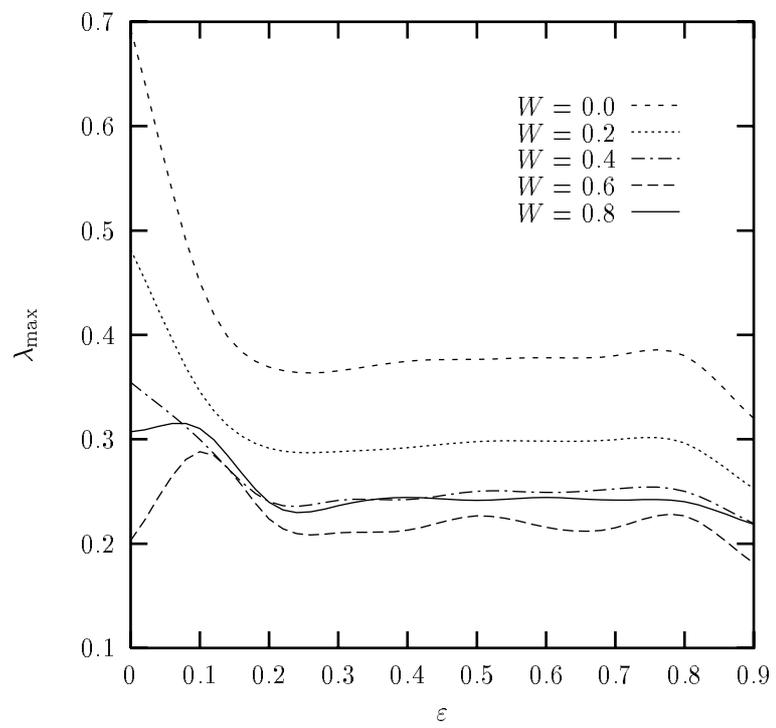
\item[Fig.~11.] The plot shows variation of $\lambda_{\rm max}$ with
coupling strength $\varepsilon$, for different values of $W$, other
system parameters remaining the same as in Fig. 1.

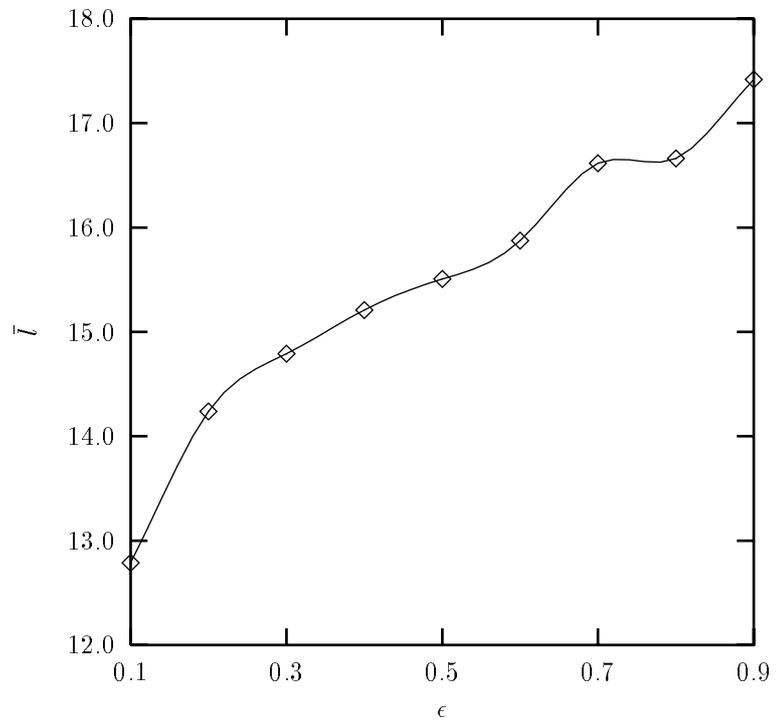
\item[Fig.~12.] Average length $\bar{l}$ of the structures is plotted
against $\varepsilon$, for $W = 0.6$.

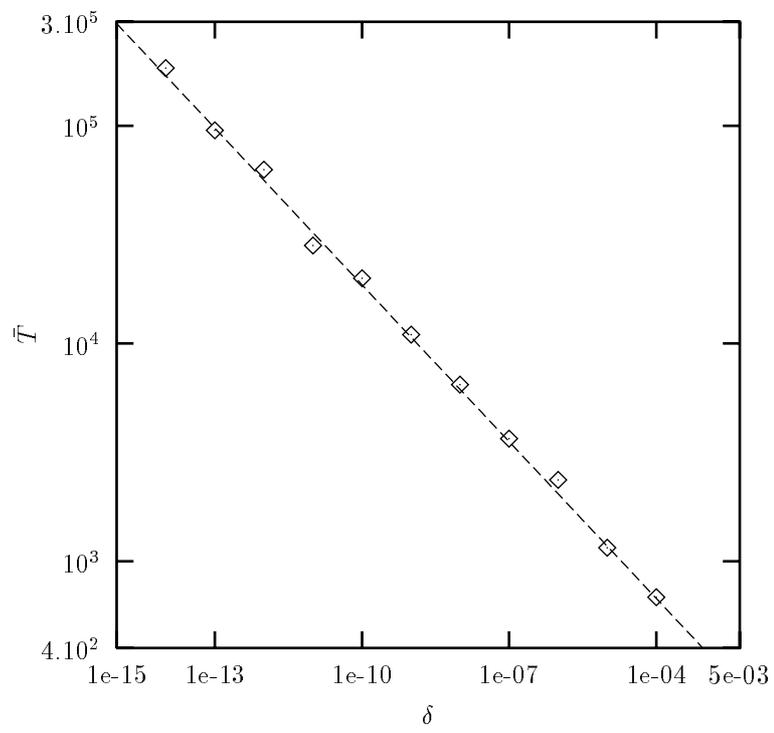
\item[Fig.~13.] Variation of $\bar{T}$ is plotted, on log--log scale,
against $\delta$, for given lattice size $L = 10000$, other parameters
being as in earlier figures.

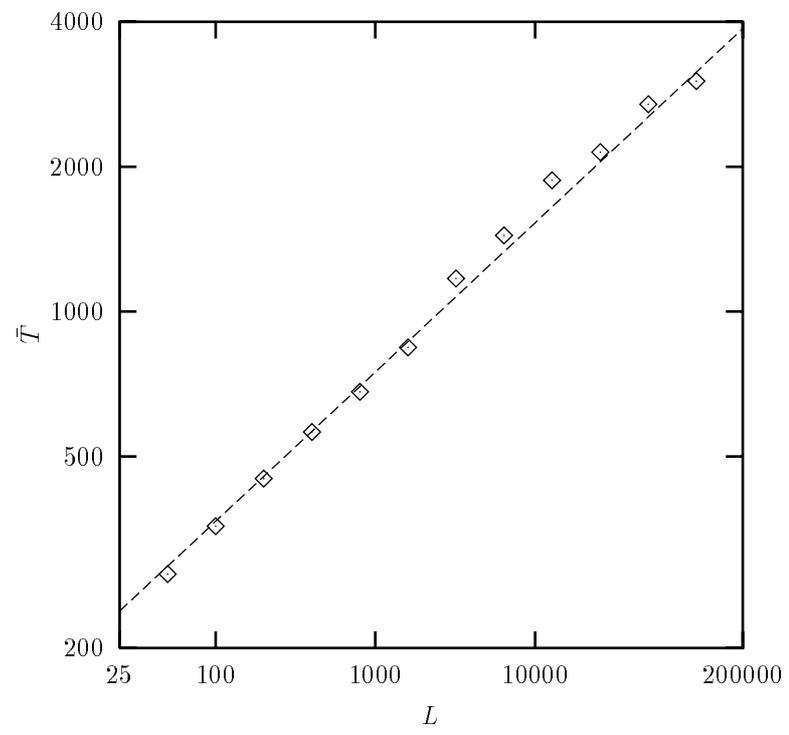
\item[Fig.~14.] The plot shows, on log--log scale, the variation of
$\bar{T}$ with lattice size $L$, for given structure parameter $\delta
= 0.0001$.

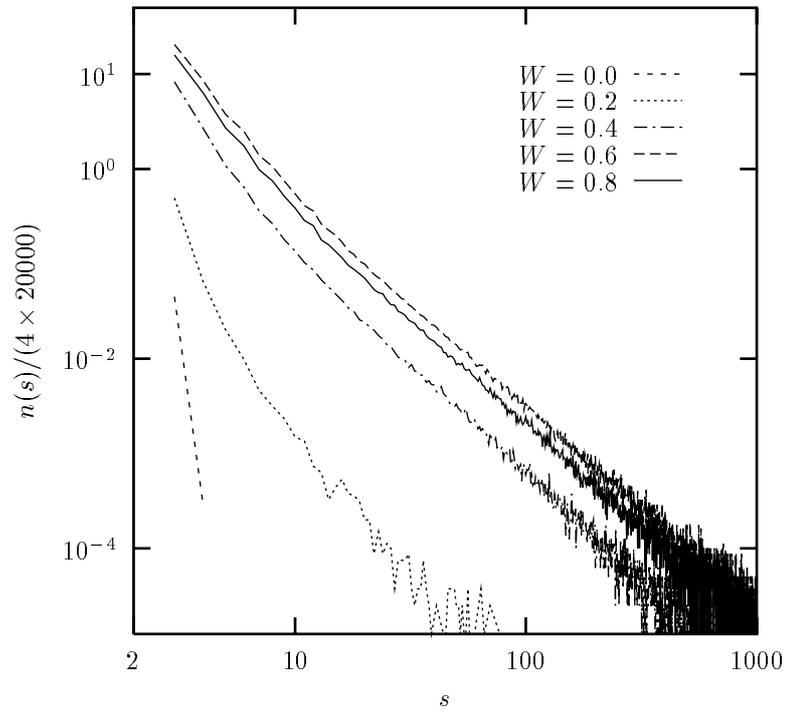
\item[Fig.~15.] Number $n(s)$ of structures is plotted, on log--log
scale, against their size
$s$, for two dimensional lattice with size $L \times L = 100 \times 100$,
for different $W$. Parameters taken are $\varepsilon = 0.6$, $\mu = 4$,
and $\delta = 0.0001$. We have used open--boundary conditions. 3000
transients are discarded. Data are obtained for 20000 iterates per
initial conditions and with 4 initial conditions.

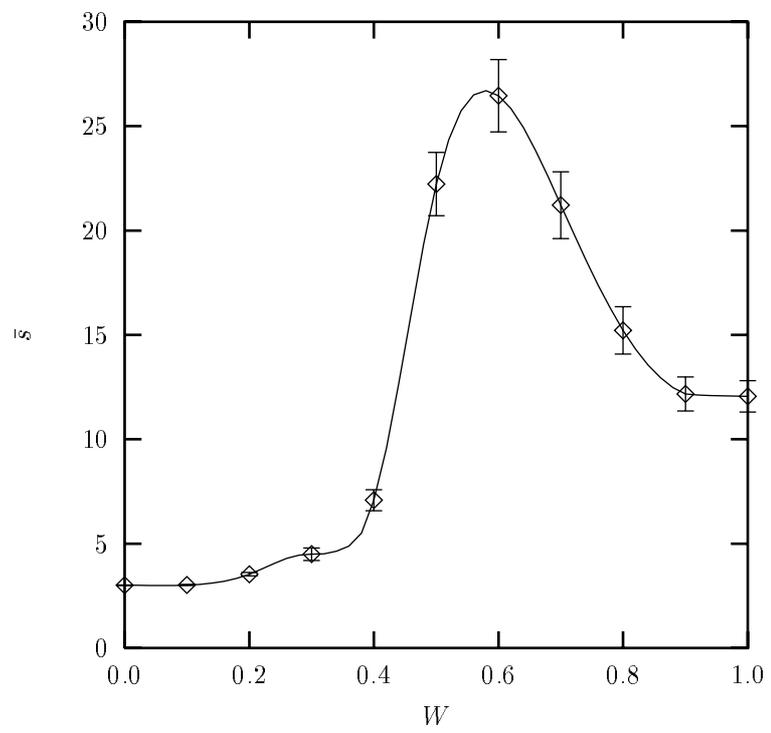
\item[Fig.~16.] The plot shows variation of average size $\bar{s}$ of the
two dimensional structures with noise $W$, for the parameters as stated
in Fig.~15. Error bars indicate the standard deviations at the data points.

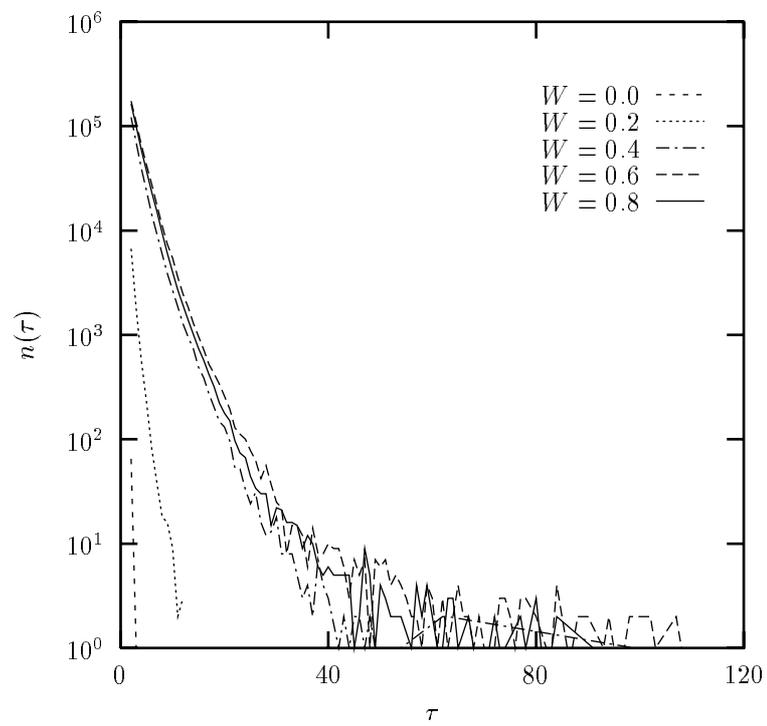
\item[Fig.~17.] Distribution of $n(\tau)$ is plotted for two dimensional
CML, on log--linear scale, against the lifetime $\tau$ of the structures,
for system parameters as in Fig.~15.

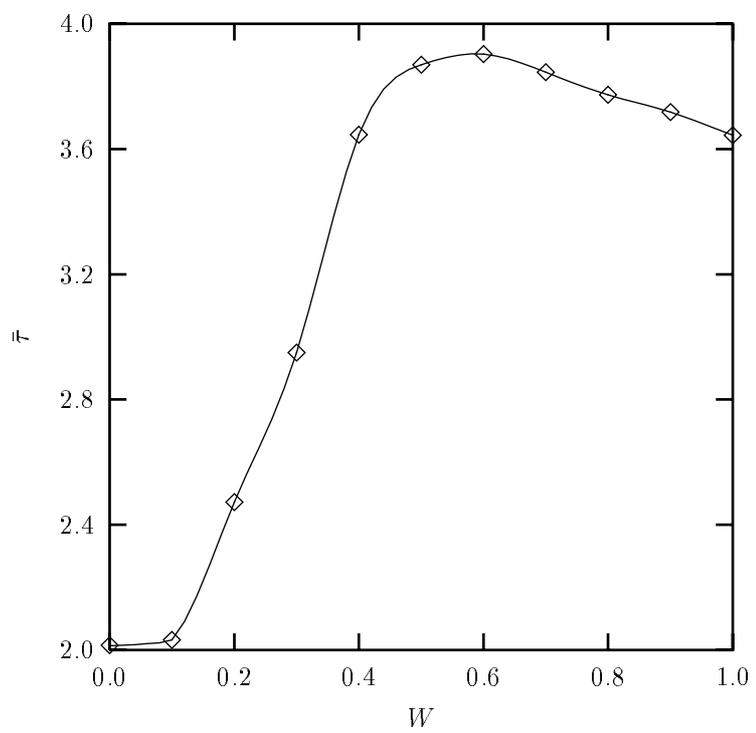
\item[Fig.~18.] Plot shows variation of $\tau$ with noise $W$, for two
dimensional CML.

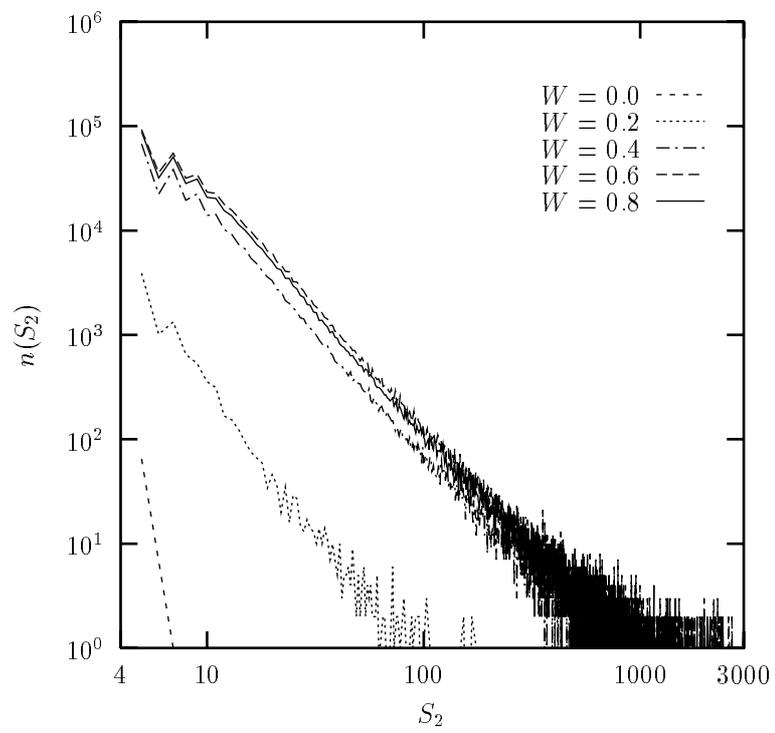
\item[Fig.~19.] Distribution of the number $n(S_2)$ of STS is plotted,
on log--log scale, against the size $S_2$, for different noise strength
values $W$, for a two dimensional CML, parameters being same as in Fig.~15.

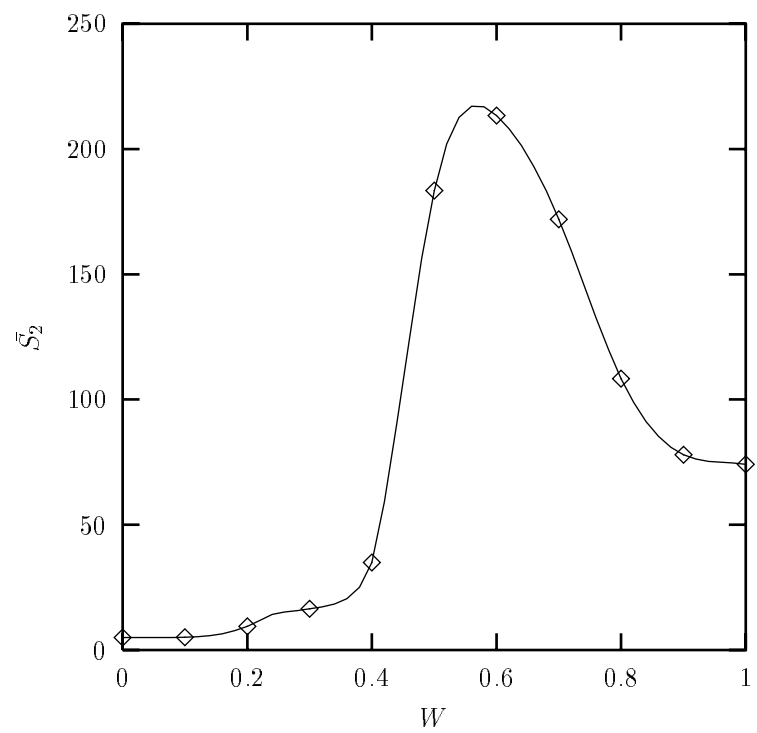
\item[Fig.~20.] Variation of average size $\bar{S_2}$ is shown against
noise strength value $W$ for a two dimensional CML.

\end{itemize}

\end{document}